\documentclass[aps,prl,twocolumn,floatfix,showpacs,superscriptaddress]{revtex4-1}

\usepackage{tcolorbox}
\usepackage{braket}
\usepackage{amsmath}
\usepackage{graphicx}
\DeclareMathOperator{\Tr}{Tr}

\begin{document}

\title{Critical slowing down in the bistable regime of circuit quantum electrodynamics} 

\author{P. Brookes}\thanks{These two authors contributed equally.}
\affiliation{Department of Physics and Astronomy, University College London, Gower Street, London, WC1E 6BT, United Kingdom}
\author{G. Tancredi}\thanks{These two authors contributed equally.}
\affiliation{Clarendon Laboratory, University of Oxford, Parks Road,
  Oxford, OX1 3PU, United Kingdom}
\author{A. D. Patterson}
\affiliation{Clarendon Laboratory, University of Oxford, Parks Road,
  Oxford, OX1 3PU, United Kingdom} 
  \author{J. Rahamim}
\affiliation{Clarendon Laboratory, University of Oxford, Parks Road,
  Oxford, OX1 3PU, United Kingdom} 
  \author{M. Esposito}
\affiliation{Clarendon Laboratory, University of Oxford, Parks Road,
  Oxford, OX1 3PU, United Kingdom} 
\author{P. J. Leek}
\affiliation{Clarendon Laboratory, University of Oxford, Parks Road,
  Oxford, OX1 3PU, United Kingdom} 
\author{E. Ginossar}
\affiliation{Advanced Technology Institute and Department of Physics,
  University of Surrey, Guildford, GU2 7XH, United Kingdom}
\author{M. H. Szymanska}
\affiliation{Department of Physics and Astronomy, University College London, Gower Street, London, WC1E 6BT, United Kingdom}

\date{\today}

\begin{abstract}
We investigate the dynamics of the bistable regime of the generalized Jaynes-Cummings Hamiltonian (GJC), realised by a circuit quantum electrodynamics (cQED) system consisting of a transmon qubit coupled to a microwave cavity. In this regime we observe critical slowing down in the approach to the steady state. By measuring the response of the cavity to a step function drive pulse we characterize this slowing down as a function of driving frequency and power. We find that the critical slowing down saturates as the driving power is increased. We compare these results with the predictions of analytical and numerical calculations both with and without the Duffing approximation. We find that the Duffing approximation incorrectly predicts that the critical slowing down timescale increases exponentially with the drive, whereas the GJC model accurately predicts the saturation seen in our data, suggesting a different process of quantum activation.
\end{abstract}

\maketitle

The study of dissipative phase transitions has a long and interesting history not only due to their technological applications, such as in the construction of the laser \cite{degiorgio1970analogy,graham1970laserlight,grossmann1971laser}, quantum limited amplifiers \cite{Siddiqi2004,Vijay2009} and optical switches \cite{Szoke1969,Gibbs1979,Amo2009}, but also due to their theoretical interest since these phase transitions cannot be described by standard techniques such as mean-field theory \cite{carmichael2015breakdown}. One of the key characteristics of dissipative phase transitions is bistability \cite{Drummond1980,mavrogordatos2017simultaneous}: close to the transition the two phases are metastable  \cite{minganti2018spectral} and the system is highly sensitive to both its parameters and its initial state \cite{reed2010high,bishop2010response,ginossar2010protocol,murch2012quantum}. The steady state is reached via rare switching events during which the system transitions from one phase to the other \cite{PhysRevA.99.043802,mavrogordatos2018rare}. This can be modelled using the theory of quantum activation in the case of dispersive optical bistability \cite{dykman_PhysRevE.75.011101}. Since the metastable states may be very long lived, this leads to critical slowing down in the equilibration time of the system. Critical slowing down has already been observed in a circuit-QED lattice \cite{Fitzpatrick2017} and an ensemble of NV centers coupled to a superconducting cavity \cite{angerer2017ultralong}, and has been modelled in the Bose-Hubbard lattice \cite{Vicentini2018a}.


In this paper, we explore the dynamics of a system consisting of a single transmon qubit coupled to a 3D superconducting microwave cavity. We observe that the transient response of the cavity exhibits critical slowing down in the bistable regime, reaching its steady state in a time much longer than the lifetimes of both qubit and cavity. By initializing the qubit in different states, we observe that this slowdown causes the cavity to retain a memory of the original qubit state throughout its transient response. Next we characterise the timescale of the slowdown as a function of driving frequency and power and we discover a new regime of behaviour at high drive powers in which the slowdown reaches a saturation that can only be explained by taking account of the quantum fluctuations of the transmon. We demonstrate this by comparing two models: the Duffing model, in which the system is treated as a single oscillator with a Kerr nonlinearity, and the generalized Jaynes-Cummings model (GJC), in which the transmon is included. At low drive powers both models are consistent with our experimental observations, since we do not expect the transmon to participate in the dynamics. However, at high drive powers the critical slowing down time reaches a saturation which is only captured by the GJC model.

\begin{figure*}
\centering
\includegraphics[width=0.9\textwidth]{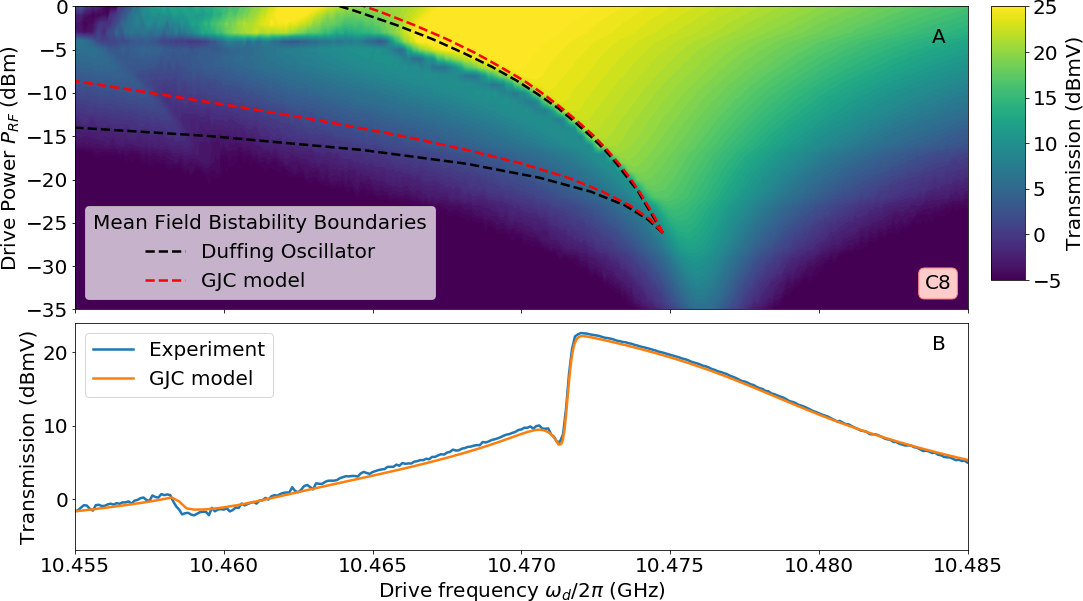}
\caption{\label{fig:spectra_contour} A: Measured signal, transmitted through the cavity, as a function of driving frequency ($\omega_d$) and power ($P_{\text{rf}}$). At low driving power, the response of the system is linear and the typical  Lorentzian lineshape is observed. As  the driving power increases, the lineshape shifts to lower frequencies and nonlinear features appear. Above $-25\,$dBm a dip in the transmitted signal is observed. This indicates the onset of the bistable regime. The boundaries of the bistable regime are calculated using mean-field theory and are indicated by red dashed lines for the GJC model and black dashed lines for the Duffing oscillator. The label C8 in the lower right hand corner indicates that this spectrum was taken during the 8th cooldown of the device. B: Signal transmitted through the cavity as a function of the driving frequency at a power  $P_{\text{rf}} = -12\,$dBm. The blue line represents the measured signal and the orange line is its simulated value attained using the master equation.}
\end{figure*}

The GJC model is defined by the following Hamiltonian:

\begin{align}
H = & \hbar \sum_n \omega_n \ket{n} \! \bra{n} + \hbar \omega_c a^\dagger a + \hbar \sum_{m,n} g_{m,n} \ket{m} \! \bra{n} (a + a^\dagger) \nonumber \\ &+ \hbar \epsilon (a^\dagger e^{- i \omega_d t} + a e^{i \omega_d t}).
\end{align}
which consists of a cavity mode of frequency $\omega_c$ coupled with strengths $g_{m,n}$ to a transmon qubit whose unperturbed eigenstates are written as $\ket{n}$ and whose energies are denoted by $\hbar \omega_n$ \cite{koch2007charge}. The cavity is represented using the annihilation(creation) operator $a$($a^\dagger$) and is driven by a monochromatic field of strength $\epsilon$ and frequency $\omega_d$.


In order to describe the effect of environmental noise on our system we make use of the Lindblad master equation \cite{Breuer2002}
\begin{align}
\label{eq:lindblad}
\partial_t \rho &= -\frac{i}{\hbar}[H,\rho] + (n_c+1)\kappa\, D(a) \rho + n_c \kappa\, D(a^\dagger) \rho \nonumber \\ 
& + \gamma_\phi D(b^\dagger b) \rho  + (n_t+1)\gamma\, D(b) \rho + n_t \gamma\, D(b^\dagger) \rho,
\end{align}
where $n_t$ and $n_c$ are the thermal occupations of the transmon and cavity bath respectively, while $\gamma$ and $\kappa$ are the intrinsic transmon and cavity relaxation rates and $\gamma_\phi$ is the intrinsic transmon dephasing rate.

In the first part of our experiment, we confirm that the GJC Hamiltonian provides a good description of our system. Using a standard cQED microwave setup, we measure the signal transmitted through the cavity as a function of driving frequency ($\omega_d$) and power ($P_{\textnormal{rf}}$) as shown in Fig. \ref{fig:spectra_contour}A. The parameters of our device are reported in Table \ref{table:parameters}. We find that at low power the cavity line is centered at $\omega_{0}/2\pi\,=\,10.4761\,$GHz and has the Lorentzian shape which is typical of linear response. As the driving power is increased, the resonance shifts towards lower frequencies and develops nonlinear features such as the dip in the transmitted signal. This dip is due to destructive interference between the two metastable states of the cavity and is a characteristic of the bistable regime \cite{Drummond1980,mavrogordatos2017simultaneous}. The two metastable states have different amplitudes and phases and are referred to as the bright and dim states according to the number of photons in the cavity. Fig.  \ref{fig:spectra_contour}B shows the measured (blue line) and the simulated (orange line) transmitted signals at $P_{\text{rf}}\,=\,-12\,$dBm. The simulated signal is obtained by  numerically solving the Lindblad Master equation (Eq. \ref{eq:lindblad}) for the steady state $\rho_{ss}$ and calculating the mean cavity amplitude tr$(\rho_{ss} a)$. We find good agreement between simulation and experiment indicating the appropriateness of our model.

 \renewcommand*{\arraystretch}{1.0}
\begin{table}
\caption{The parameters used to model the device during the 8th cooldown (C8) are listed below. \label{table:parameters}}
\begin{ruledtabular}
\begin{tabular}{p{0.2in}p{0.6in}p{0.2in}p{0.6in}}
\multicolumn{2}{c}{Hamiltonian Parameters} & \multicolumn{2}{c}{Lindblad Parameters} \\
\hline
$\omega_c / 2 \pi$ & $10.423\,$GHz & $\kappa / 2 \pi$ & $1.432\,$MHz \\
$E_J / 2 \pi$ & $46.7\,$GHz & $\gamma / 2 \pi$ & $33\,$kHz \\
$E_C / 2 \pi$ & $221\,$MHz & $\gamma_\phi / 2 \pi$ & $1\,$kHz \\
$g_0 / 2 \pi$ & $295\,$MHz & $n_c$ & $0.01$ \\
 $ \Delta_0 / 2 \pi $ & $ -1.572\,$GHz & $n_t$ & 0.02 \\
\end{tabular}
\end{ruledtabular}
\end{table}

\begin{figure}
\centering
\includegraphics[width=0.45\textwidth]{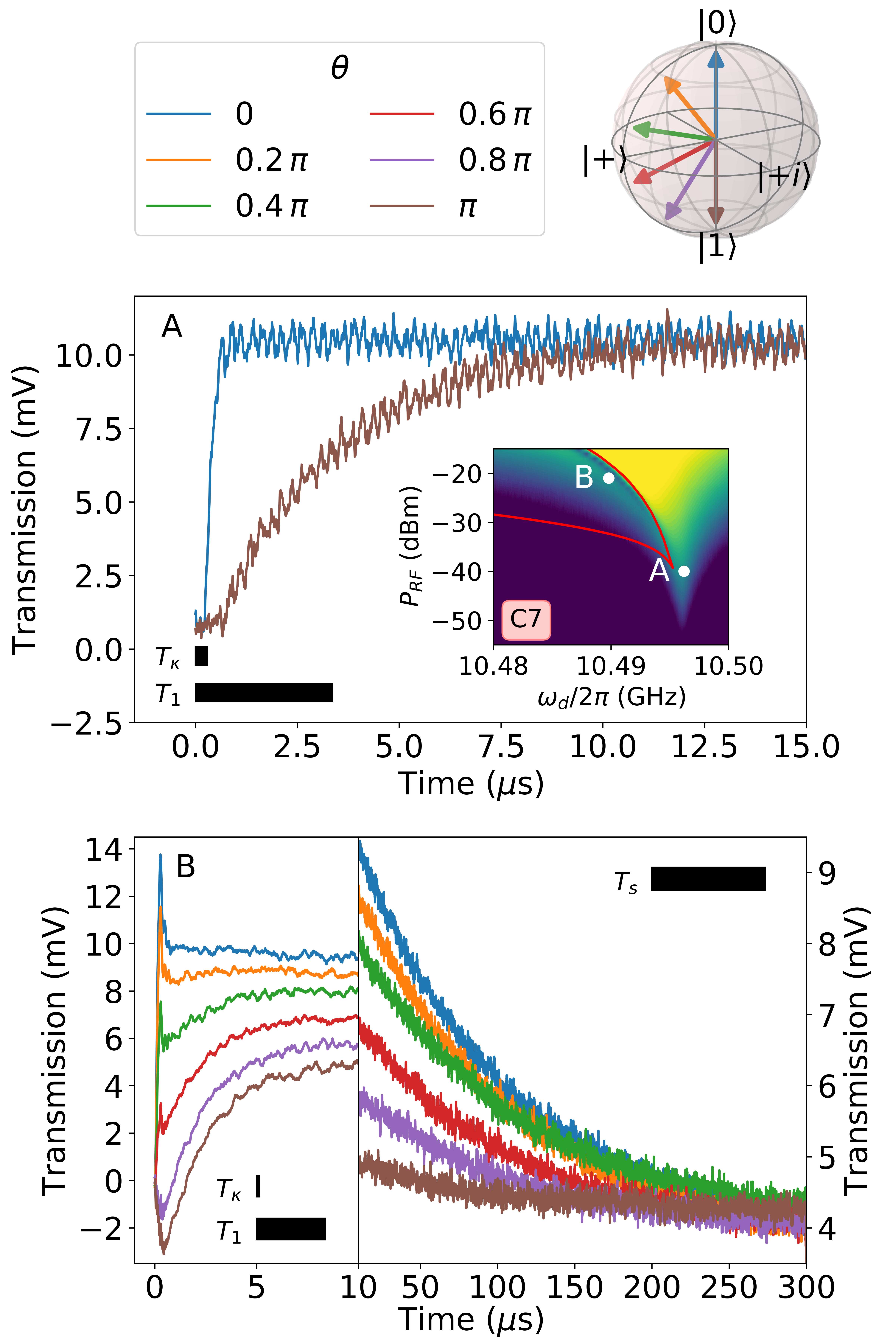}
\caption{\label{fig:memory} Averaged transient response of the cavity outside and inside the bistable regime. The inset in panel A shows the mean-field GJC limits of the bistable regime in red and indicates the locations at which the data in panels A and B were taken. (A): The cavity is driven at the low power resonance $\omega_d / 2 \pi = 10.4671\,$GHz and $P_{\text{rf}}=-40\,$dBm. The signal in blue (brown) is the transient response measured with the qubit initialized in its ground (first excited) state. The transient response is governed by the timescale $T_\kappa=0.29\mu$s if the transmon is in the ground state, corresponding to $\theta = 0$ on the Bloch sphere, whereas it is governed by $T_1=3.37\mu$s if the transmon is in the first excited state ($\theta = \pi$). (B): Transient responses for different initial qubit states in the bistable regime at $P_{\text{rf}}=-21\,$dBm. The transient response is divided into two parts. There is an initial fast response with a time scale ranging from $T_\kappa$ to $T_1$ depending on the initial transmon state, followed by a slow decay towards steady state over a timescale $T_s=73.2\mu\text{s}$ which is much longer than both the transmon and cavity lifetimes. This critical slowing down allows to distinguish the transients for different transmon states for over $100\mu$s.}
\end{figure}

We also model the boundaries of the bistable regime using mean-field theory, in which the state of the system is approximated by a product of coherent states $ \ket{\psi} = \ket{\alpha} \otimes \ket{\beta} $. By substituting this into Eq. (\ref{eq:lindblad}) and solving for $ \partial_t \alpha = 0 $ and $ \partial_t \beta = 0$, we identify the region of parameter space which produces two stable fixed points (Appendix A). The boundaries of the GJC  bistable regime are shown by the red dashed lines in Fig. \ref{fig:spectra_contour}A.  We also apply the Duffing approximation \cite{mavrogordatos2017simultaneous}, in which the system is approximated to a single oscillator with a Kerr nonlinearity $K$,
\begin{equation}
    \widetilde{H} = \hbar \widetilde{\omega}_c a^\dagger a + \frac{1}{2} \hbar K a^\dagger a^\dagger a a + \hbar \widetilde{\epsilon} (a^\dagger e^{- i \omega_d t} + a e^{i \omega_d t}),
\end{equation}
which interacts with its environment via the Lindblad operators $\sqrt{(1+\widetilde{n}_c) \widetilde{\kappa}} a$, $\sqrt{\widetilde{n}_c \widetilde{\kappa}} a^\dagger$ and $\sqrt{\widetilde{\kappa}_\phi} a^\dagger a$. This is explained in detail in Appendix B. The resulting boundaries are shown by the black dashed lines. The GJC model and the Duffing approximation produce results which are qualitatively similar: the bistable regime emerges just below the resonance frequency at a drive power of $P_{\text{rf}}=-27\,$dBm and opens up over an increasingly wide range of frequencies as the drive power is increased. Although the lower bound on the drive frequency of the bistable regime differs, they appear to have a similar behaviour for this range of driving powers.

We now focus on understanding the system dynamics by measuring the transient response of the cavity when a step function drive pulse is applied. Fig. \ref{fig:memory} shows the average cavity response outside (A) and inside (B) the bistable regime. The response of the cavity at the low power resonance with the transmon initialized in either the ground state (blue line) or the first excited state (brown line)  are shown in Fig. \ref{fig:memory}A. The timescale over which the cavity responds shows a clear dependence on the transmon state. When the transmon starts in the ground state the cavity reaches equilibrium over a timescale set by the cavity relaxation rate $\kappa$, $T_\kappa = 2 \pi / \kappa\,=\,0.29\,\mu$s; whereas, when the transmon is initialized in the first excited state, the drive is initially off resonant with the cavity and we must wait for the transmon to relax over a time $T_1\,=\,3.37\,\mu$s before the cavity can reach equilibrium.
 
 \begin{figure}[h]
\centering
\includegraphics[width=0.4\textwidth]{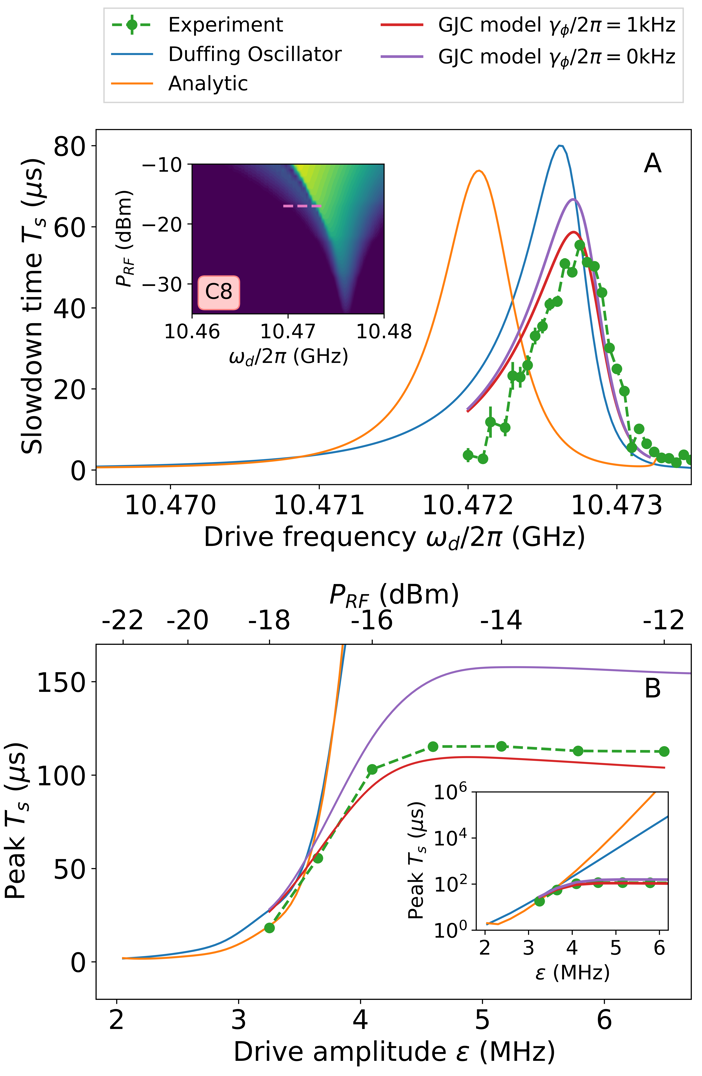}
\caption{\label{fig:dykman_slowdown} (A): Critical slowing down time $T_s$ in the bistable regime as a function of driving frequency at $P_{\text{rf}} = -17$dBm. The green points represent the experimental data, which we compare with the results of master equation calculations applied to the Duffing oscillator (blue line) and the GJC model with transmon dephasing (red line) and without (purple line). We also display the results of previous analytical theory of switching rates for the Duffing oscillator (orange line) \cite{dykman_PhysRevE.75.011101}. At this power both the master equation and the analytical calculation qualitatively reproduce the experimental values of $T_s$. The pink line in the inset shows the location of our measurements within the cavity spectrum. (B): Maximum value of $T_s$ for different drive amplitudes. As the drive power increases beyond $-17$dBm, $T_s$ reaches a saturation at a value of $\approx 100\,\mu$s, that is consistent with the simulations based on the GJC model with transmon dephasing (red line). Removing the dephasing by setting $\gamma_\phi = 0$ (purple line) does not change the power at which saturation occurs but it does raise the upper limit on $T_s$. Meanwhile analytical (orange line) and master equation (blue line) calculations with the Duffing approximation predict that $T_s$ rises exponentially with drive amplitude, as can be seen using the logarithmic scale of the inset.}
\end{figure}
 
The dynamics are completely different when the system is in the bistable regime, as shown in Fig. \ref{fig:memory}B. The cavity response is now governed by two  different timescales. First, there is a fast rise in the cavity transmitted signal over a time ranging from $T_{\kappa}$ to $T_1$ depending on the initial state of the qubit. Then we observe critical slowing down, a gradual decay towards equilibrium over a time much longer than both the cavity and qubit lifetimes. We label the time constant over which the system reaches equilibrium as $T_s$. By initializing the transmon in a range of initial states, we see that the cavity retains a memory of the initial transmon state for over $100 \mu$s.

In order to model the critical slowing down, we define the occupational probabilities for the bright and dim states as $p_b$ and $p_d$ and write a simple rate equation:
\begin{equation}\label{eq:rate_eq}
\frac{d}{dt} \begin{pmatrix} 
p_b \\ 
p_d
\end{pmatrix}
=
\begin{pmatrix} 
\Gamma_{d \rightarrow b} &  -\Gamma_{b \rightarrow d} \\ 
-\Gamma_{d \rightarrow b} & \Gamma_{b \rightarrow d}
\end{pmatrix}
\begin{pmatrix} 
p_b \\ 
p_d
\end{pmatrix}.
\end{equation}
This model is a valid description of the occupation probabilities once the system has entered one of the two metastable states. At this point, the system will proceed to make random jumps, governed by Poissonian statistics, from bright to dim at a rate $\Gamma_{b\rightarrow d}$ and from dim to bright at a rate $\Gamma_{d \rightarrow b}$. The solution to this model can be easily written down as
\begin{equation}\label{eq:rate_sol}
\begin{pmatrix} 
p_b \\ 
p_d
\end{pmatrix}
=
\frac{1}{\Gamma_{b \rightarrow d} + \Gamma_{d \rightarrow b}}
\begin{pmatrix}
\Gamma_{b \rightarrow d} \\ 
\Gamma_{d \rightarrow b}
\end{pmatrix}
+ A e^{-t/T_s}
\begin{pmatrix}
1 \\ 
-1
\end{pmatrix},
\end{equation}
where the critical slowing down time is given by
\begin{equation}
T_{s} = \frac{1}{\Gamma_{d \rightarrow b} + \Gamma_{b \rightarrow d}}.
\end{equation}

This means that the critical slowing down time is entirely determined by the switching rates between the metastable states. We use several different methods to model these rates and display the results in Fig. \ref{fig:dykman_slowdown} together with experimental results (green points). Firstly, we apply the Duffing approximation and calculate the switching rates, and hence $T_s$, using the analytical theory of quantum activation provided in \cite{dykman_PhysRevE.75.011101} (orange line). Secondly, we estimate  $T_s$ by calculating the asymptotic decay rate of the Lindblad master equation of the Duffing oscillator (blue line). Finally we find the asymptotic decay rate of the Lindblad master equation describing the full GJC model of our device both with transmon dephasing (red line) and without (purple line). Fig. \ref{fig:dykman_slowdown}A shows the variation of $T_s$ with drive frequency $\omega_d$ along the dashed line in the spectroscopy inset, which is located at a drive power of $P_{\text{rf}}=-17\,$dBm that covers the range of bistability predicted by the mean-field calculations displayed in Fig. \ref{fig:spectra_contour}. The results show a peak in $T_s$, which is centered on the dip in the transmission. This is qualitatively reproduced by the Duffing model.

However, if we plot how the maximum value of $\,T_s$ varies with the amplitude of the drive, Fig. \ref{fig:dykman_slowdown}B, we observe a significant difference between our data and the simulated values attained with the two theoretical methods which model the system as a single Duffing oscillator. Whereas the theory of the Duffing oscillator predicts that $T_s$ should increases exponentially with the drive, we instead observe that, at sufficiently strong drive amplitudes, $T_s$ saturates. To explain this discrepancy we require the full GJC model. When we explicitly include the transmon in the simulation we find that the master equation predicts the same ceiling in $T_s$ as is found in experiment.

Clearly the system of two strongly coupled oscillators is governed by essentially different activation dynamics. In the Duffing oscillator the mean number of photons always increases with drive amplitude along with the critical slowing down time. However, here we note that the saturation in the critical slowing down time occurs simultaneously with a saturation in the mean number of excitations in the transmon (Appendix C). This saturation keeps the system in a regime where the fluctuations in the number of excitations in the transmon is dominant over the mean-field values. This distinguishes the two-oscillator system from the Duffing oscillator and may be the reason for the very different dependence on parameters observed in the critical slowing down time.

In summary, we observe critical slowing down in a system consisting of a single transmon qubit coupled to a 3D cavity. We find that this critical slowing down is well modelled by the Duffing approximation at low drive powers, whereas at high drive powers we observe a saturation in the critical slowing down time, which can only be captured by the full GJC model. It is known that in this regime the transmon becomes more highly excited and starts to participate in the dynamics so it is no longer valid to apply the Duffing approximation \cite{mavrogordatos2017simultaneous}. An accurate model must include the quantum fluctuations of the qubit. Currently there exists no analytical theory for the switching rates in the bistable regime of a cavity coupled to spins or multilevel systems and this suggests one avenue of future work could focus on extending the existing theory for the Duffing oscillator to these models.

Furthermore, when measuring the transient response of the cavity, the observed critical slowing down allows us to distinguish different initial states of the transmon over a timescale far longer than the lifetimes of both the cavity and the transmon. With the use of a quantum limited amplifier it should be possible to identify which metastable state the system is occupying at any point in time during a single experimental shot, thus, allowing the measurement of the occupation probabilities of the bright and dim states for different initial transmon states, potentially leading to a novel readout mechanism.

This work has received funding from the EPSRC under grant nos. EP/J001821/1, EP/J01350/1, EP/M013243/1 and EP/K003623/2. The data underlying this work is available without restriction.

\onecolumngrid
\appendix

\section{Supplemental Information}

\subsection{A: Obtaining the mean field equations of motion}
The equations of motion for the cavity and and transmon amplitudes are given by:
\begin{align}
\partial_t \alpha &= \Tr \big(a \partial_t \rho \big) \\
\partial_t \beta &= \Tr \big(b \partial_t \rho \big). 
\end{align}
Using the Eq. (2) from the main text this can be rewritten as:
\begin{align}
\partial_t \alpha =& -i \Tr \big( [H, a\,] \rho \big) - \frac{1}{2} \kappa\, \Tr \big( a \rho \big) \label{eq:eom_alpha} \\
\partial_t \beta =& -i \Tr \big( [H, b \,] \rho \big) - \frac{1}{2} (\gamma + \gamma_\phi) \Tr \big( b \rho \big). \label{eq:eom_beta}
\end{align}
The mean field theory of our system is obtained by approximation the state of the system as a product of coherent states:
\begin{equation}
\label{eq:mean_field_approximation}
\rho \approx \ket{\alpha} \! \bra{\alpha} \otimes \ket{\beta} \! \bra{\beta}.
\end{equation}
In order to express the equations of motion as polynomial functions of $\alpha$ and $\beta$ we must first express the Hamiltonian as a normal ordered product of creation and annihilation operators. We start by considering an operator $O$ acting on a single mode:
\begin{equation}
    O = \sum_{x,y=0}^{\infty} C_{x,y} a^{\dagger x} a^y.
\end{equation}
The question we must answer is: how can we calculate the coefficients $C_{x,y}$ which produce $O$? The matrix elements of $O$ are given by:
\begin{equation}
    \bra{x} O \ket{y} = \sum_{k=0}^{\textnormal{min}(x,y)}\frac{\sqrt{x!y!}}{k!}C_{x-k,y-k}.
\end{equation}
Let us take $O$ to be a transition operator $O=\ket{m}\!\bra{n}$. In this case we find
\begin{equation}
\label{eq:transition_matrix_elements}
    \delta_{x,m} \delta_{y,n} = \sum_{k=0}^{\textnormal{min}(x,y)}\frac{\sqrt{x!y!}}{k!}C_{x-k,y-k}.
\end{equation}
This equation can be used to inductively demonstrate that $C_{x,y}=0$ if $x-y \neq m-n$. To do this it is convenient to change the indices of $C$ so that they are measured relative to $m$ and $n$. We can do this by writing $ x = m + \chi + \Delta $ and $ y = n + \chi - \Delta $. If we define the quantity
\begin{equation}
\label{eq:sum}
    G_{\chi,\Delta,m,n} = \sum_{k=0}^{\textnormal{min}(m+\chi+\Delta,n+\chi-\Delta)} \frac{C_{m+\chi+\Delta-k,n+\chi-\Delta-k}}{k!}
\end{equation}
then we can use Eq. (\ref{eq:transition_matrix_elements}) to obtain
\begin{align}
\label{eq:sum_result}
\ G_{\chi,\Delta,m,n}  = \frac{\delta_{\Delta}\delta_{\chi}}{\sqrt{(m+\chi+\Delta)!(n+\chi-\Delta)!}}.
\end{align}
The condition $x-y \neq m-n$ is equivalent to $\Delta \neq 0$, in which case the sum above always vanishes. If we then take $\chi=\chi_{\textnormal{min}}=-\textnormal{min}(m+\Delta,n-\Delta)$ the sum contains only a single term at $k=0$ which can only be zero if $C_{m+\chi_{\textnormal{min}}+\Delta,n+\chi_{\textnormal{min}}-\Delta}=0$. Increasing $\chi$ by $1$ will introduce an additional term to the sum, but since the previous term is already known to be zero and their sum is known to be zero the new term must also be zero. By incrementally increasing $\chi$ we can show that $C_{m+\Delta-\chi,n-\Delta-\chi} =0$ for all $\chi$ provided that $\Delta \neq 0$.

The more interesting case arises when $\Delta=0$. We can repeat the previous argument to prove $C_{m+\chi,n+\chi}=0$ for $\chi < 0$ but at $\chi=0$ the summation $G_{0,0,m,n}$ does not vanish. Instead we use Eq. (\ref{eq:sum_result}) to find
\begin{equation}
    C_{m,n}= \frac{1}{\sqrt{m!n!}}.
\end{equation}
For $\chi>0$ the situation is more complicated because although the sum vanishes again it now contains multiple non-zero terms. Fortunately $C_{m+\chi,n+\chi}$ can be expressed in terms of $C_{m,n}$ as follows. Let us rewrite Eq. (\ref{eq:sum}) by reindexing the sum by $d=\chi-k$, making use of $C_{m+\chi,n+\chi}=0$ for $\chi < 0$ and suppressing the $m$, $n$ and $\Delta$ indices:
\begin{equation}
    G_{\chi} = \sum_{d=0}^{\chi}\frac{C_{m+d,n+d}}{(\chi-d)!}.
\end{equation}
Using the fact that $G_\chi=0$ for $\chi>0$ we can write
\begin{align}
0 &= \sum_{f=0}^{\chi-1} \frac{(-1)^f}{f!} G_{\chi-f} \\
&=\sum_{f=0}^{\chi-1} \sum_{d=0}^{\chi-f} \frac{(-1)^f}{f!} \frac{C_{m+d,n+d}}{(\chi-d-f)!} \\
&= \frac{(-1)^{\chi+1}}{\chi!} C_{m,n} + \sum_{d=0}^{\chi} \sum_{f=0}^{\chi-d} \frac{(-1)^f}{f!} \label{eq:sum_of_sums} \frac{C_{m+d,n+d}}{(\chi-d-f)!}.
\end{align}
This can be simplified using the binomial distribution:
\begin{equation}
    (u+v)^N = \sum_{r=0}^{N} \frac{N!}{(N-r)!r!} u^{r} v^{N-r}.
\end{equation}
Taking $u=-1$, $v=1$ and $N \geq 1$ the sum above is clearly zero and we obtain
\begin{equation}
   \sum_{r=0}^{N} \frac{(-1)^{r}}{(N-r)!r!} = 0.
\end{equation}
Using this in Eq. (\ref{eq:sum_of_sums}) we find
\begin{equation}
    C_{m+d,n+d} = \frac{(-1)^{\chi}}{\chi!} C_{m,n}.
\end{equation}
So the transition matrix $\ket{m}\!\bra{n}$ can be expressed as
\begin{equation}
    \ket{m}\!\bra{n} = \frac{1}{\sqrt{m!n!}} \sum_{\chi \geq 0} \frac{(-1)^\chi}{\chi!} (a^\dagger)^{m+\chi} a^{n+\chi}.
\end{equation}
The Hamiltonian then becomes
\begin{align}
H = & \,\hbar \omega_c a^\dagger a + \hbar \sum_{n=0}^{\infty} \sum_{\chi \geq 0} \frac{\omega_n(-1)^\chi}{n!\chi!} (b^\dagger)^{n+\chi} b^{n+\chi} \nonumber \\ &+ \hbar (a + a^\dagger) \sum_{m,n=0}^{\infty} \sum_{\chi \geq 0} \frac{g_{m,n}}{\sqrt{m!n!}} \frac{(-1)^\chi}{\chi!} (b^\dagger)^{m+\chi} b^{n+\chi} \nonumber \\ 
&+ \hbar \epsilon (a^\dagger e^{- i \omega_d t} + a e^{i \omega_d t}).
\end{align}
The commutators are
\begin{align}
    [H,a\,] &= - \hbar \omega_c a -\hbar \epsilon e^{i \omega_d t} \nonumber \\ 
   &- \hbar \sum_{m,n=0}^{\infty} \sum_{\chi \geq 0} \frac{g_{m,n}}{\sqrt{m!n!}} \frac{(-1)^\chi}{\chi!} (b^\dagger)^{m+\chi} b^{n+\chi} \\
    [H,b\,] &= -\hbar \sum_{n=0}^{\infty} \sum_{\chi \geq 0} \frac{(n+\chi)\omega_n(-1)^\chi}{n!\chi!} (b^\dagger)^{n+\chi-1} b^{n+\chi} \nonumber \\
    & - \hbar (a + a^\dagger) \sum_{m,n=0}^{\infty} \sum_{\chi \geq 0} \frac{(-1)^\chi (m+\chi)g_{m,n}}{\chi! \sqrt{m!n!}} (b^\dagger)^{m+\chi-1} b^{n+\chi} .
\end{align}
Using the above results in combination with the mean-field approximation in Eq. (\ref{eq:mean_field_approximation}) and the equations of motion given by Eqs. (\ref{eq:eom_alpha},\ref{eq:eom_beta}) we find
\begin{align}
    \partial_t \alpha &= i \hbar \omega_c \alpha + i \hbar \epsilon e^{i \omega_d t}  - \frac{1}{2}\kappa \alpha \nonumber \\ 
   & + i \hbar \sum_{m,n=0}^{\infty} \sum_{\chi \geq 0} \frac{g_{m,n}}{\sqrt{m!n!}} \frac{(-1)^\chi}{\chi!} (\beta^*)^{m+\chi} \beta^{n+\chi} \\
   \partial_t \beta &=  i \hbar \sum_{n=0}^{\infty} \sum_{\chi \geq 0} \frac{(n+\chi)\omega_n(-1)^\chi}{n!\chi!} (\beta^*)^{n+\chi-1} \beta^{n+\chi} \nonumber \\
    & +i \hbar (\alpha + \alpha^*) \sum_{m,n=0}^{\infty} \sum_{\chi \geq 0} \frac{(-1)^\chi (m+\chi)g_{m,n}}{\chi! \sqrt{m!n!}} (\beta^*)^{m+\chi-1} \beta^{n+\chi} \nonumber \\
    &- \frac{1}{2} ( \gamma + \gamma_\phi) \beta.
    \end{align}
We can find the steady state by solving these equations for $\partial_t \alpha = 0$ and $\partial_t \beta = 0$. In the bistable regime there will be two stable solutions to these equations, while outside there will be only one.

\subsection{B: Performing the Duffing approximation}

In the main text we approximate our system as a Duffing oscillator in order to obtain a benchmark showing the dynamics of a single non-linear oscillator in the bistable regime. This model is used to calculate the boundaries of the bistable regime in the mean-field approximation, as in Fig. 1, and to model how the critical slowing down time varies with drive amplitude, as in Fig. 3. To map our system to a Duffing oscillator we project the GJC Hamiltonian onto a low energy subspace and identify a Kerr nonlinearity in the resulting spectrum. This subspace consists of the eigenstates of the GJC Hamiltonian for which the transmon is close to the ground state. The first step is to identify these states. We start by writing the the Hamiltonian in the form:
\begin{equation}
    H = H_0 + H_{\text{int}}
\end{equation}
where $H_0$ describes the bare cavity and transmon, and $H_{\text{int}}$ describes the interaction between them. These components are given by \cite{koch2007charge}:
\begin{align}
    H_0 &= \hbar \omega_c a^\dagger a + \hbar \sum_{n=0}^{\infty} \omega_n \ket{n} \! \bra{n} \\
    H_{\text{int}} &= \hbar \sum_{n=0}^{\infty}g_n \big( a \ket{n+1}\!\bra{n} + a^\dagger \ket{n} \! \bra{n+1} \big).
\end{align}
If the interaction is turned off by setting $g_n = 0$ then the eigenstates of $H$ are simply products of the eigenstates of the bare cavity and transmon with eigenstates and eigenenergies given by:
\begin{align}
H_{0}\ket{m}\ket{n} &= E_{mn} \ket{m} \ket{n} \\
   E_{mn} &= \hbar(m \omega_c + \omega_n)
\end{align}
where $m$ denotes the number of photons in the cavity and $n$ denotes the number of excitations in the transmon. For finite strength interactions we enter the dispersive regime, which is defined by $\lvert g_n / \Delta_n \rvert \ll 1$ where the detuning is given by
\begin{equation}
    \Delta_n = \omega_{n+1} - \omega_n - \omega_c.
\end{equation}
Provided that the interaction strength is sufficiently weak we can continue to label the eigenstates by the number of cavity and transmon excitations they carry and if the system is weakly driven close to the cavity resonance then the only state which will take part in the dynamics are those for which the transmon is in the dressed ground state. These states form a ladder of dressed cavity states which define the low energy subspace upon which we can project our model. We define the projector by:
\begin{equation}
    \Pi = \big(\ket{\psi_{0,0}}\!\bra{\psi_{0,0}},\ket{\psi_{1,0}}\!\bra{\psi_{1,0}},\ket{\psi_{2,0}}\!\bra{\psi_{2,0}},... \big)
\end{equation}
where $\ket{\psi_{mn}}$ represents the eigenstate of $H$ which can be smoothly transformed to $\ket{m}\ket{n}$ by turning off $H_{\text{int}}$. Using this projector we obtain the low energy model:
\begin{align}
    \widetilde{H} &= \Pi^\dagger H \Pi \\
    &= \hbar \widetilde{\omega}_c a_0^\dagger a_0 + \frac{1}{2} \hbar K a_0^\dagger a_0^\dagger a_0 a_0 + \mathcal{O} \big( (g_m/\Delta_m)^6 \big).
\end{align}
where the ladder operator in the projected subspace is defined by
\begin{equation}
    a_0 = \sum_{n=1}^{\infty} \sqrt{n} \ket{\psi_{n-1,0}} \! \bra{\psi_{n,0}}.
\end{equation}
This Hamiltonian describes a Duffing oscillator with a frequency $\widetilde{\omega}_c$ and a Kerr nonlinearity $K$ \cite{murch2012quantum}. If we use the GJC model parameters given in Table 1 of the main text then the Duffing model parameters we obtain are shown in Table S1 below. This table also includes the rescaled drive amplitude $\widetilde{\epsilon}$ which arises when we add a driving term to $H_0$ of the form:
\begin{equation}
    H_d(t) = \hbar \epsilon (a^\dagger e^{- i \omega_d t} + a e^{i \omega_d t}).
\end{equation}
We find that this transforms to a similar driving term in the Duffing Hamiltonian given by
\begin{equation}
    \widetilde{H}_d(t) = \hbar \widetilde{\epsilon} (a_0^\dagger e^{- i \omega_d t} + a_0 e^{i \omega_d t}).
\end{equation}
The drive amplitude in the projected space $\widetilde{\epsilon}$ is given by $r_a \epsilon$ where $r_a$ is calculated according to: 
\begin{align}
    \Pi^\dagger \, a \, \Pi &= r_a \, a_0 + \mathcal{O}\big( (g_m / \Delta_m)^4 \big)
\end{align}
Next we must consider the Lindblad operators which describe the effects of environmental interactions. We have already considered $a$, but the remaining operators can be projected into the low energy subspace as follows:
\begin{align}
    \Pi^\dagger \, b \, \Pi &= r_b \, a_0 + \mathcal{O}\big( (g_m / \Delta_m)^2 \big) \\
    \Pi^\dagger \, b^\dagger b \, \Pi &= r_{nb} a_0^\dagger a_0 + \mathcal{O}\big( (g_m / \Delta_m)^2 \big).
\end{align}
The coefficients of the Lindblad operators in the low energy subspace are then given by
\begin{align}
    a_0 :& \,\,\,\, \sqrt{(1+\widetilde{n}_c) \widetilde{\kappa}} = \sqrt{r_a^2 (1+n_c) \kappa + r_b^2 (1+n_t) \gamma} \\
    a_0^\dagger :&  \,\,\,\, \sqrt{\widetilde{n}_c \widetilde{\kappa}} = \sqrt{r_a^2 n_c \kappa + r_b^2 n_t \gamma} \\
    a_0^\dagger a_0 :& \,\,\,\, \sqrt{\widetilde{\kappa}_\phi} = \sqrt{r_{nb}^2 \gamma_\phi} 
\end{align}
The full set of Lindblad parameters in the Duffing model are given in Table S1.

 \renewcommand*{\arraystretch}{1.0}
\begin{table}
\renewcommand\thetable{S1} 
\caption{When we project the GJC model of our system onto the Duffing model we obtain the parameters of the Duffing model are listed below. \label{table:parameters}}
\begin{ruledtabular}
\begin{tabular}{p{0.2in}p{0.6in}p{0.2in}p{0.6in}}
\multicolumn{2}{c}{Hamiltonian Parameters} & \multicolumn{2}{c}{Lindblad Parameters} \\
\hline
$\widetilde{\omega}_c / 2 \pi$ & $10.4761\,$GHz & $\widetilde{\kappa} / 2 \pi$ & $1.387\,$MHz \\
$K / 2 \pi$ & $-0.152\,$MHz & $\widetilde{\kappa}_\phi / 2 \pi$ & $1.02\,$Hz \\
$\widetilde{\epsilon}$ & $ 0.984 \epsilon$ & $\widetilde{n}_c$ & 0.0100 \\
\end{tabular}
\end{ruledtabular}
\end{table}

\subsection{C: Saturation in the number of transmon excitations}

\begin{figure}[h]
\centering
\includegraphics[width=0.6\textwidth]{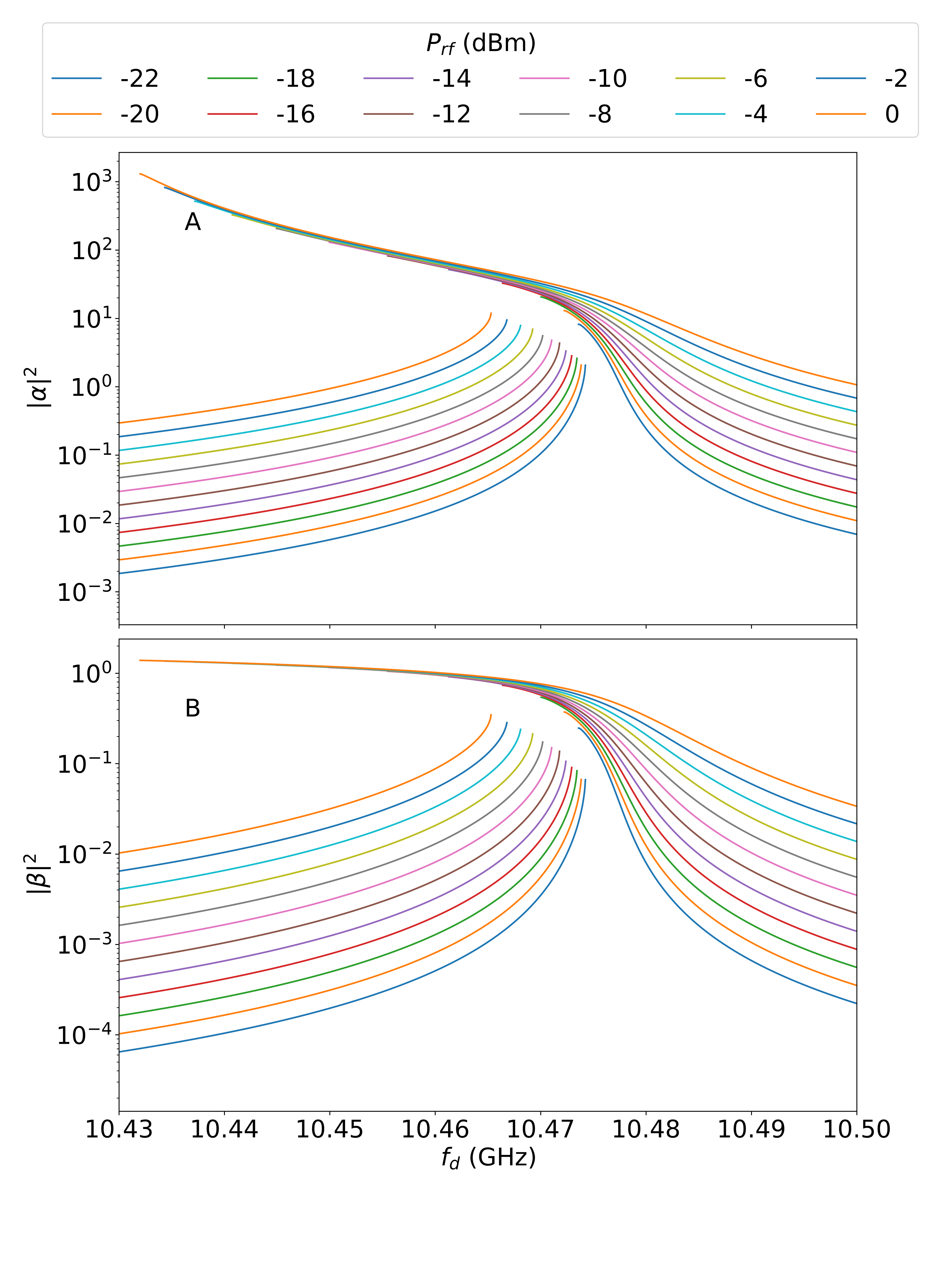}
\renewcommand\thetable{FIG. S1.} 
\caption{\label{fig:gjc_mf} Above we plot the mean field number of photons in the cavity (panel A) and number of excitations in the transmon (panel B) for the bright and dim metastable states against the frequency of the drive. We can see that as the drive power is increased from $-22$dBm to $0$dBm the maximum number of photons in the bright state increases by two orders of magnitude up to $\sim 1000$ photons. Meanwhile the transmon saturates and reaches only $\sim 1$ excitation.}
\end{figure}

In the main text we plot the maximum value of the slowdown time $T_s$ as it varies with drive amplitude, and we find that it reaches a saturation in the strongly driven limit for both the experimental data and the GJC model. Meanwhile the Duffing model predicts that $T_s$ continues to grow exponentially. The key distinction between the GJC model and the Duffing model is the inclusion of the transmon, which finds itself restricted to low lying excitations even when the cavity is strongly driven. This restricts the transmon to a regime in which we expect quantum fluctuations to play a significant role in the dynamics.

We can see this by observing the results shown in in Fig. \ref{fig:gjc_mf}. Here we apply the mean-field approximation and plot the mean occupations of the cavity and the transmon in the steady state, given by $\Tr(a^\dagger a \rho) = \lvert \alpha \rvert^2$ and $\Tr(b^\dagger b \rho) = \lvert \beta \rvert^2$ respectively. For each drive power there are two branches of solutions: one corresponding to the bright state and the other corresponding to the dim state. We can see that in the bright state the cavity reaches $\sim1000$ photons at a drive of $P_{\text{rf}}=0$dBm while the transmon still only has $\sim1$ excitation.

\begin{figure}[h]
\centering
\includegraphics[width=0.7\textwidth]{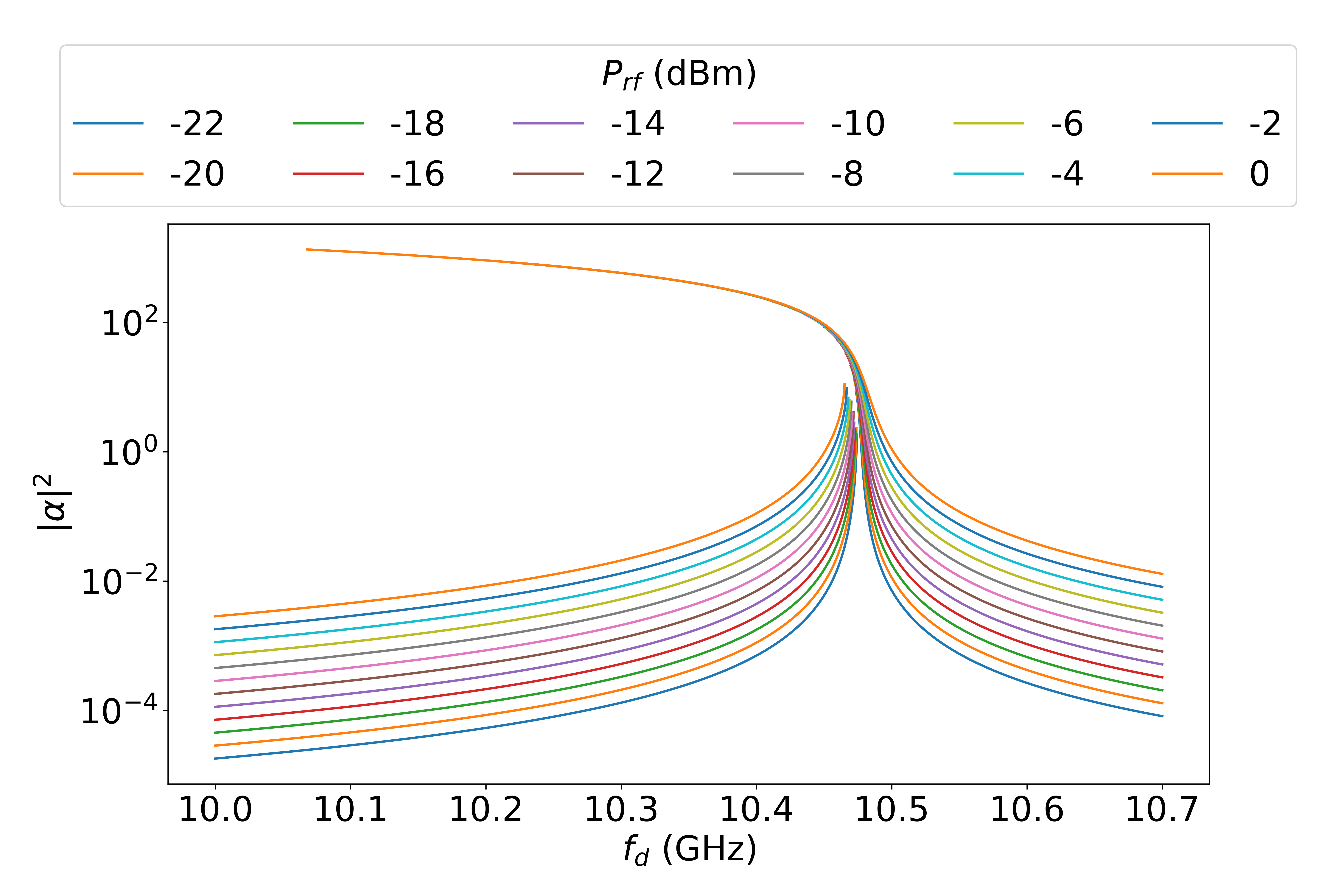}
\renewcommand\thetable{FIG. S2.} 
\caption{\label{fig:duffing_mf} Here we plot the mean field number of photons for the bright and dim metastable states of the Duffing oscillator against the frequency of the drive. When the drive power is increased from $-22$dBm to $0$dBm the maximum number of photons in the bright state increases up to $\sim 1000$ photons.}
\end{figure}

For comparison we also plot the mean occupation of the Duffing oscillator in Fig. \ref{fig:duffing_mf}. We see that the number of photons in the cavity reaches $\sim1000$ for $P_{\text{rf}}=0$dBm.

\bibliography{Bibliography}

\end{document}